\newif\ifARXIV
\ARXIVtrue

\ifARXIV
\else
\fi

\documentclass[conference]{IEEEtran}

\ifCLASSOPTIONcompsoc
  \usepackage[nocompress]{cite}
\else
  \usepackage{cite}
\fi

\usepackage{amsfonts}  

\usepackage{amsmath}  

\usepackage{multirow}

\usepackage{hyperref}

\usepackage{graphicx}

\usepackage{listings}

\usepackage{caption}
\begin{document}

\title{Performance Overhead of Atomic Crosschain Transactions}

\author{
    \IEEEauthorblockN{Peter Robinson}
    \IEEEauthorblockA{Protocol Engineering Group and Systems, ConsenSys}
    \IEEEauthorblockA{School of Information Technology and Electrical Engineering, University of Queensland, Australia\\
    peter.robinson@uqconnect.edu.au}
}

\ifARXIV
\else
\IEEEoverridecommandlockouts
\IEEEpubid{\makebox[\columnwidth]{978-1-7281-8086-1/20/\$31.00~\copyright2020 IEEE \hfill} \hspace{\columnsep}\makebox[\columnwidth]{ }}
\fi

\maketitle

\ifARXIV
\thispagestyle{plain}
\pagestyle{plain}
\else
\IEEEpubidadjcol
\pagestyle{plain}
\fi

\begin{abstract}
Atomic Crosschain Transaction technology allows composable programming across permissioned Ethereum blockchains. 
It allows for inter-contract and inter-blockchain function calls that are both synchronous and atomic: if one part fails, the whole call graph of function calls is rolled back. 
This paper analyses the processing overhead of using this technique compared to using multiple standard non-atomic single blockchain transactions. 
The additional processing is analysed for three scenarios involving multiple blockchains: the Hotel - Train problem, Supply Chain with Provenance, and an Oracle. 
The technology is shown to reduce the performance of Hyperledger Besu from 375 tps to 39.5 tps if all transactions are instigated on one node, or approaching 65.2 tps if the transactions are instigated on a variety of nodes, for the Hotel-Train scenario.
\end{abstract}


\section{Introduction}
\label{sec:introduction}
Atomic Crosschain Transactions\cite{crosschainwhitepaper, robinson2019b} is an Ethereum permissioned blockchain technology\cite{robinson2018a, enteth20} that allows transactions across blockchains that either update state on all blockchains or discarded state updates on all blockchains. These crosschain transactions are nested transactions that consist of an Originating Transaction, Subordinate Transactions and Subordinate Views. Originating and Subordinate Transactions update the state of the blockchains they run on. Subordinate Views return the result of a function call on one blockchain to another blockchain. The Atomic Crosschain Transactions are able to provide this functionality because they provide cross-blockchain consensus.

Providing cross-blockchain consensus does not come for free. BLS Threshold Signatures \cite{bls2004,bls-threshold,bls-threshold-youtube} are used to prove values across blockchains. Verifying BLS Threshold Signatures uses two costly BLS Pairing operations. Generating BLS Threshold Signatures involves communications between validator nodes in a blockchain network and verifying the signature prior to publication. Validating information from other blockchains requires signature verification. Additionally, the Coordination Contract on a Coordination Blockchain, used to coordinate parts of the overall transaction, needs to verify signatures. The contribution of this paper is the first analysis of the additional processing requirements imposed by using the Atomic Crosschain Transaction technology compared to using separate single blockchain transactions. 

The Atomic Crosschain Transaction technology can be used to create applications to solve a wide variety of problems. Three common example problems involving transactions reading and updating across multiple blockchains are presented so that the technology can be evaluated. The \textit{Hotel Train} problem has a travel agency book a hotel room and a train seat. This requires the state of three blockchains to be updated atomically. The \textit{Supply Chain Provenance} problem allows for selective transparency of supply chain events between a Supply Chain blockchain and a Provenance blockchain. This requires two blockchains to update atomically. The \textit{Oracle} problem involves using the result of a function call on one blockchain, for example to read the current price of a commodity, in a function call on another blockchain. 

Creating blockchain systems that provide atomic behaviour across multiple blockchains is a complex problem\cite{crosschain-deals}. This is because blockchain networks are distributed systems and cross-blockchain protocols thus operate in a distributed system of distributed systems. To date, Atomic Crosschain Transactions is the only protocol that provides a true atomic cross-blockchain function call protocol \cite{robinson-consensus-crosschain}. Understanding the overhead imposed by a cross blockchain protocol is important as this allows application architects to evaluate whether the additional utility provided by the protocol outweighs the overhead.

This paper is organised as follows: the \textit{Background} section introduces Ethereum. Next three example problems are described in the \textit{Scenarios} section. The Atomic Crosschain Transactions technology is described in the \textit{Framework} section. The \textit{Experimental Setup} section describes how throughput numbers were gathered. The \textit{Results} section analyses the overhead of using Atomic Crosschain Transaction technology compared to using separate single blockchain transactions. The affect of Byzantine actors is described in the final section.

\section{Background}
\label{ref:ethereum}
Ethereum\cite{wood2016a} is a blockchain platform that allows users to deploy and execute computer programs known as Smart Contracts. 
Ethereum transactions update the state of the distributed ledger, do not return values, and can emit log information. They can be used to deploy contracts, execute function calls and transfer Ether, the native coin of Ethereum, between accounts. ``View" function calls can be executed on the Smart Contract code. These View function calls return a value and do not update the state of the Smart Contract. 

Blocks consist of groups of transactions. Ethereum nodes come to consensus to determine the next block that will become part of the chain of blocks. As such, execution of transactions involves nodes across the network. In contrast, View function calls execute on a single node using the node's local copy of the distributed ledger. 

Ethereum MainNet, the most widely used public Ethereum network, uses a Proof of Work (PoW) \cite{nakamoto2008, wood2016a} consensus algorithm. Ethereum 2, due to go live incrementally starting in 2020, will use a Proof of Stake (PoS) \cite{ethereum2-beacon-chain} consensus algorithm. Ethereum is also deployed in permissioned consortium networks\cite{enteth20}. In these deployments Proof of Authority (PoA) consensus algorithms such as Istanbul Fault Byzantine Tolerant version 2 (IBFT2)\cite{ibft2} are used. 


\section{Scenarios}
\label{sec:scenarios}
\subsection{Hotel and Train}
A common example distributed transaction problem is know as the Hotel and Train problem. In this scenario, a travel agent needs to ensure the atomicity of a combined booking transaction. In other words, the travel agent needs to ensure that they either book both the hotel room and the train seat, or neither, so that they avoid the situation where a hotel room is successfully booked but the train reservation fails, or vice versa. There are three permissioned blockchains involved: the travel agency runs a blockchain, and each hotel and train travel company also maintains its own blockchain. 

\begin{figure} [h]     
\includegraphics[width=\columnwidth]{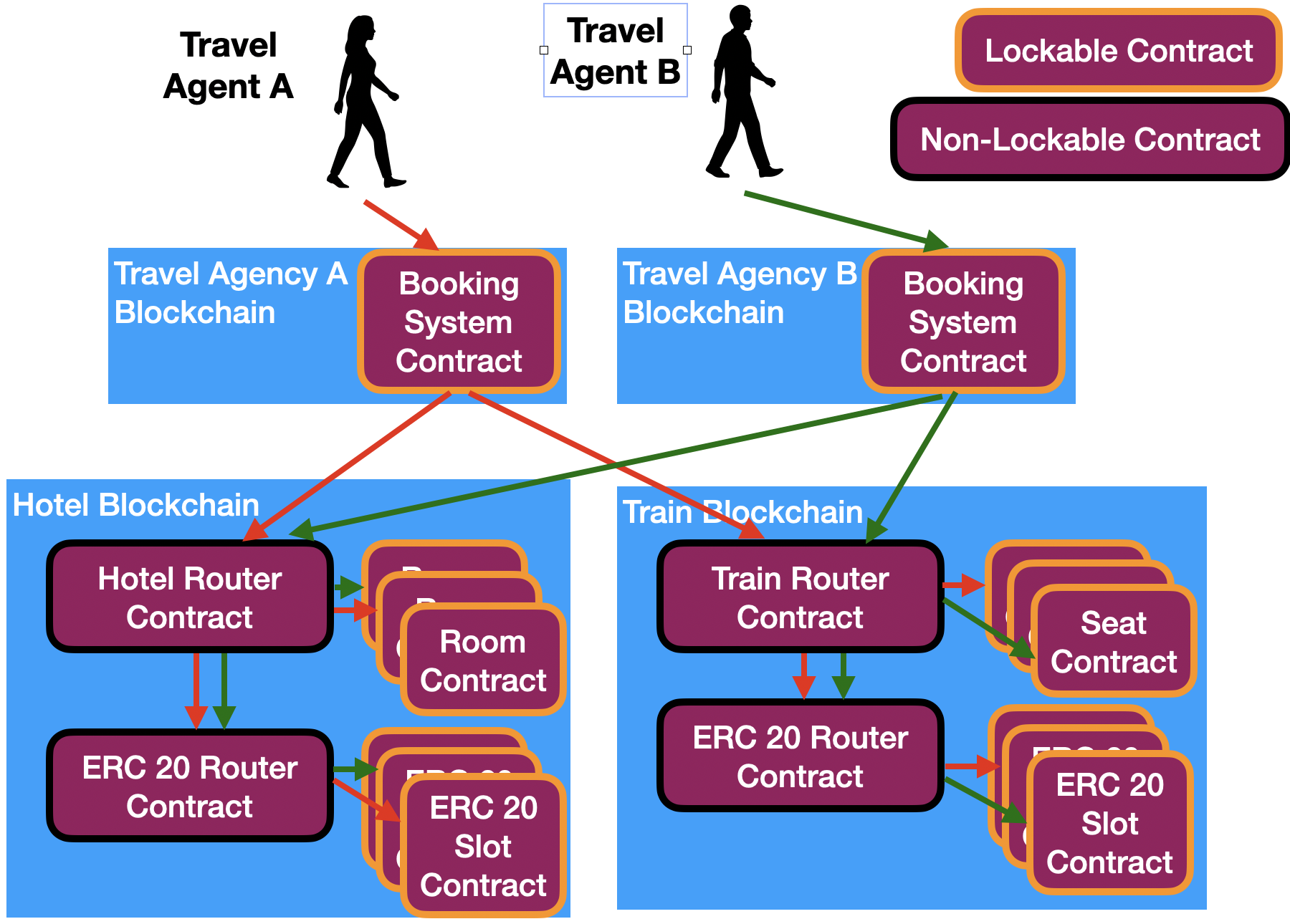}       
\caption{Hotel Train Scenario}
\label{fig:hoteltrain}
\end{figure}

Figure~\ref{fig:hoteltrain} illustrates the implementation of a `Hotel and Train' reservation system using Atomic Crosschain Transaction technology. The train blockchain and the hotel blockchain host especially designed contracts that conform to a Router-Item pattern. With this pattern, a non-lockable router contract is used to access lockable item contracts. For example, when booking a train seat, a Train Router contract function is called that locates a Train Seat contract that is not locked that indicates it represents a train seat that is available for the date that the train seat needs to be reserved for. In this way multiple parallel cross blockchain calls that need to lock contracts that contain updates can occur in parallel, thus allowing for multiple seat reservations to occur in parallel. Similarly, the ERC 20 token contracts 
\cite{eip20} have been modified to conform to this pattern, which allows for multiple parallel payments. 

Travel agents purchase ERC 20 tokens on the hotel and train blockchains, which they can then use to pay for accommodation and travel. They book hotel rooms and train seats by executing cross blockchain transactions, paying for the rooms and seats in the same transaction.

\subsection{Supply Chain with Provenance}
A vendor may wish to publish information to customers to provide assurances of product provenance. With only a single global blockchain all transaction details must be published and will be visible to all participants of the global blockchain. The vendor may prefer, for example, to keep the identities of its suppliers secret from competing vendors. 

\begin{figure} [h]     
\includegraphics[width=\columnwidth]{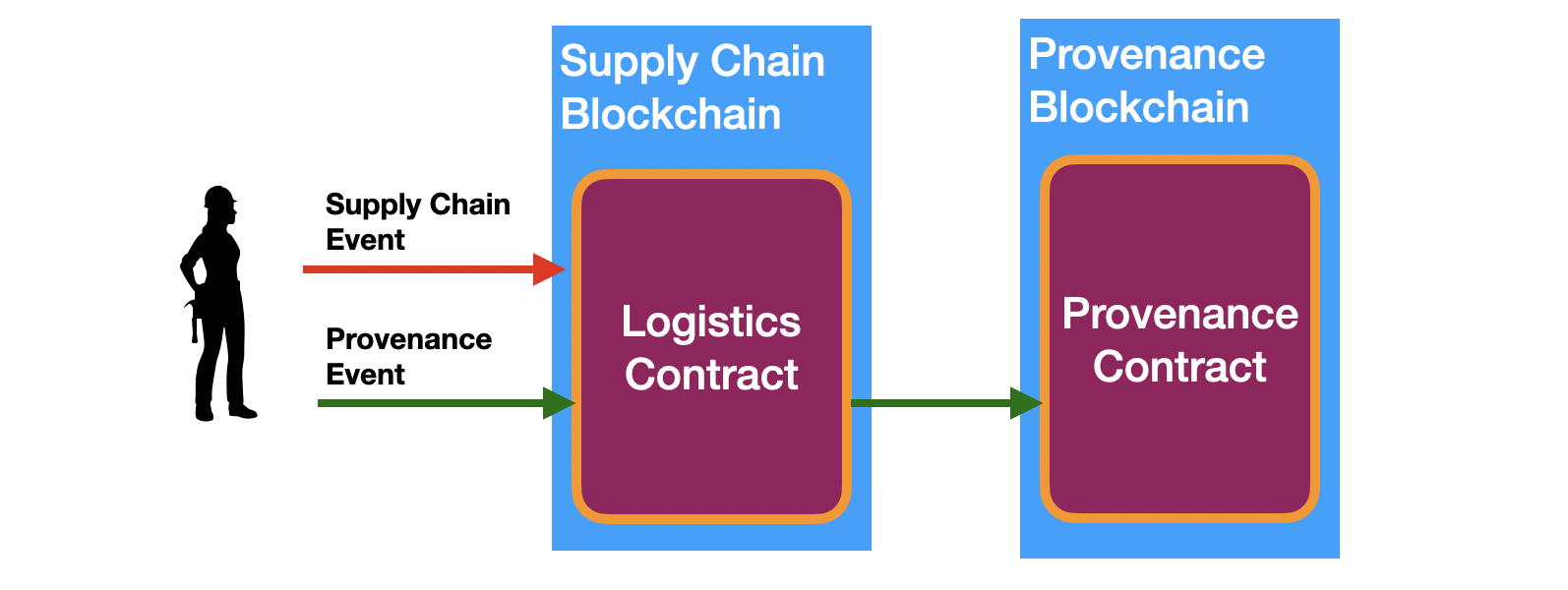}
\caption{Supply Chain with Provenance Scenario}
\label{fig:supplychain}
\end{figure}

This scenario is illustrated in Figure.~\ref{fig:supplychain} in which two blockchains are used. A \textit{Supply Chain} blockchain maintains all of the transactions between a vendor and its suppliers. The \textit{Provenance} blockchain holds all the information required to assure customers of various aspects of the goods being purchased. When a provenance event occurs, a cross blockchain transactions is used to update the Supply Chain blockchain and the Provenance blockchain. 

\subsection{Oracle}
\begin{figure} [h]     
\includegraphics[width=\columnwidth]{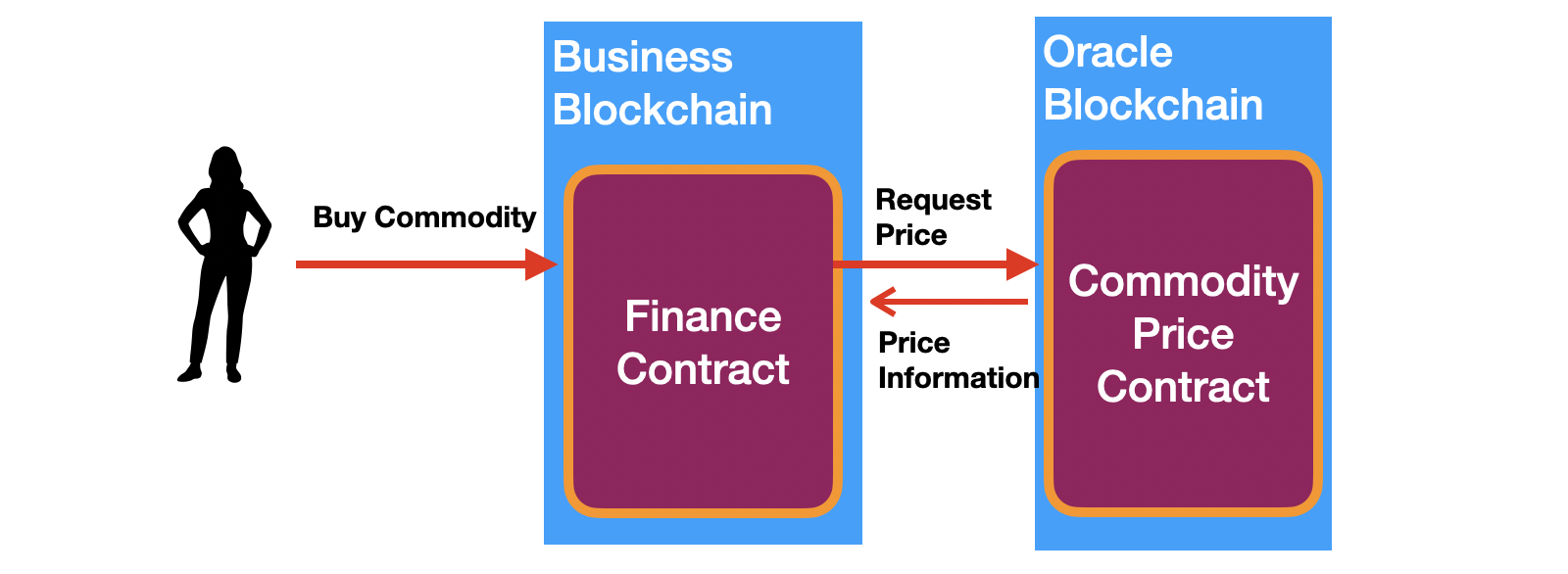}
\caption{Oracle Scenario}
\label{fig:oracle}
\end{figure}

An Oracle blockchain maintains a set of data that is valuable to other blockchain applications. For example, in Fig.~\ref{fig:oracle} a company could publish commodity prices to their blockchain. They could charge other companies for the right to access their blockchain. The other companies could use cross blockchain transactions to execute business logic on their own blockchain based on the information returned from the Oracle blockchain. 

\section{Framework}
\label{ref:atomic}
Atomic Crosschain Transaction technology\cite{crosschainwhitepaper, robinson2019b} has been designed to shield application developers from the complexity of crosschain transactions by incorporating the required changes into the Ethereum Client software. The technology has been implemented in a fork of Hyperledger Besu\cite{besu}, and is available on github.com\cite{crosschain-github}. 

\subsection{Nested Transactions}
\begin{figure}
  \includegraphics[width=\linewidth]{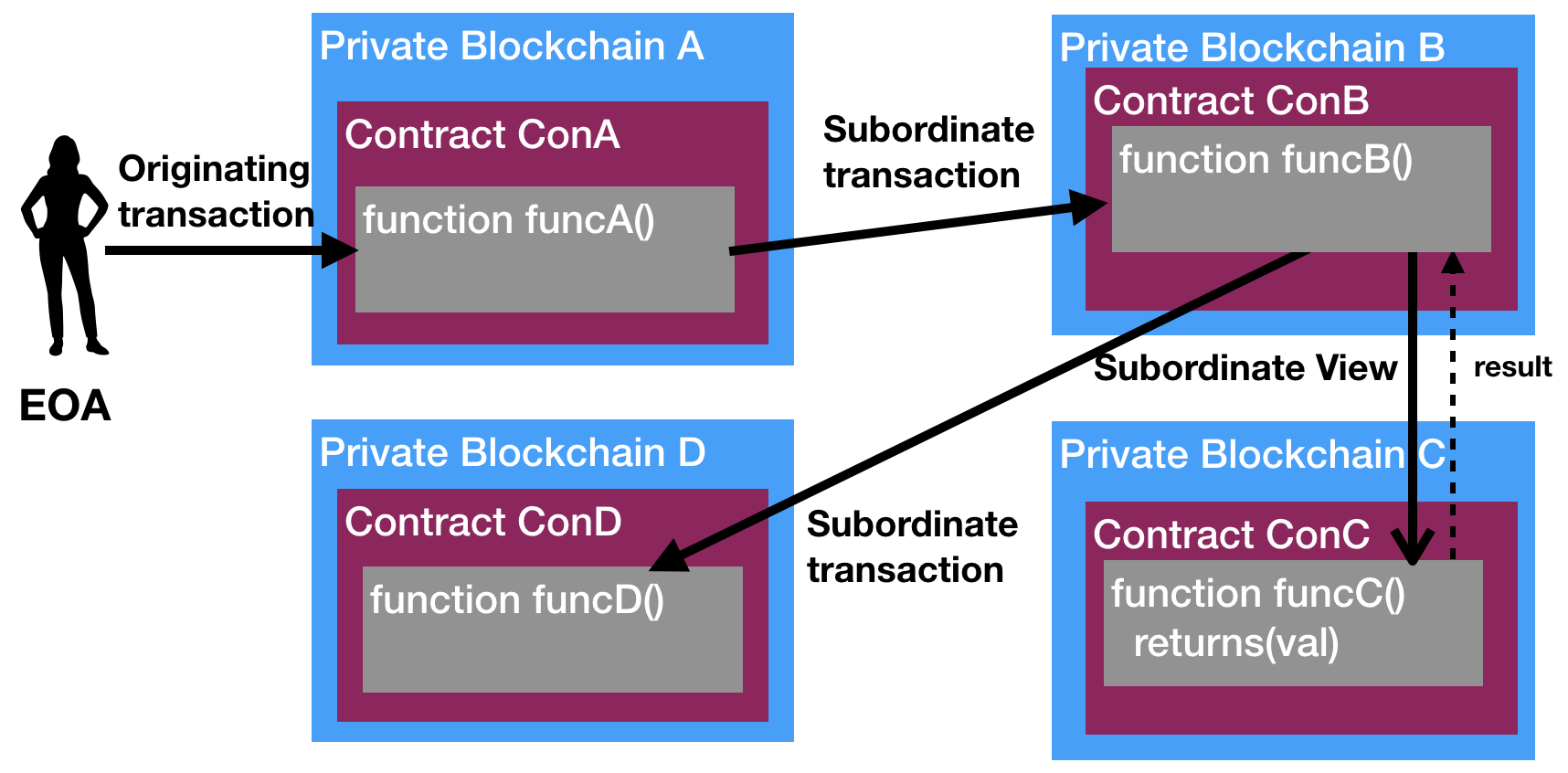}
  \caption{Originating Transaction containing Two Nested Subordinate Transactions and a Subordinate View}
  \label{fig:nested1}
\end{figure}

Atomic Crosschain Transactions are nested Ethereum transactions and views. Figure~\ref{fig:nested1} shows an Externally Owned Account (EOA) calling a function \texttt{funcA} in contract \texttt{ConA} on blockchain \texttt{Private Blockchain A}. This function in turn calls function \texttt{funcB}, that in turn calls functions \texttt{funcC} and \texttt{funcD}, each on separate blockchains. The transaction submitted by the EOA is called the \textit{Originating Transaction}. The transactions that the Originating Transaction causes to be submitted are called Subordinate Transactions. Subordinate Views may also be triggered. In Figure~\ref{fig:nested1}, a Subordinate View is used to call \texttt{funcC}. This function returns a value to \texttt{funcB}.

The EOA user constructs the nested transaction by first creating the signed Subordinate View for \texttt{Private Blockchain C} and the signed Subordinate Transaction for \texttt{Private Blockchain D}. They then create the signed Subordinate Transaction for \texttt{Private Blockchain B}, encapsulating the signed Subordinate Transaction and View. Finally, they sign the Originating Transaction for \texttt{Private Blockchain A}, including the signed Subordinate Transactions and View.  

\subsection{Per-Node Transaction Processing}
When the EOA submits the Originating Transaction to a node, the node processes the transaction using the algorithm shown in Listing~\ref{listing:processing}. If the transaction includes any Subordinate Views, they are dispatched and their results are cached (Lines 1 to 3). The function is then executed (Lines 4 to 17). If a Subordinate Transaction function call is encountered, the node checks that the parameter values passed to the Subordinate Transaction function call match the parameter values in the signed Subordinate Transaction (Lines 6 to 8). If a Subordinate View function call is encountered, the node checks that the parameters passed to the Subordinate View function call match the parameter values in the signed Subordinate View (Lines 9 and 10). The cached values of the results of the Subordinate View function calls are then returned to the executing code (Line 11). If the execution has completed without error, then each of the signed Subordinate Transactions is submitted to a node on the appropriate blockchain (Nodes 18 to 20).

\begin{lstlisting}[
%  frame=single,
  basicstyle=\footnotesize\ttfamily,
  numbers=left,
stepnumber=1, 
  firstnumber=1,
  numberfirstline=true,
  numbersep=5pt,    
  xleftmargin=0.5cm,
  morekeywords={msg},
  label=listing:processing,
  caption=Originating or Subordinate Transaction Processing
]
For All Subordinate Views {
  Dispatch Subordinate Views & cache results
}
Trial Execution of Function Call {
  While Executing Code {
    If Subordinate Transaction function called {
      check expected & actual parameters match.
    } 
    Else If Subordinate View function is called {
      check expected & actual parameters match
      return cached results to code
    } 
    Else {
      Execute Code As Usual
    }
  }
}
For All Subordinate Transactions {
  Submit Subordinate Transactions
}
\end{lstlisting}

\subsection{Blockchain Signing and Threshold Signatures}
BLS Threshold Signatures \cite{bls-threshold, bls-threshold-youtube} combines the ideas of threshold cryptography \cite{shamir1979} with Boneh-Lynn-Shacham(BLS) signatures \cite{bls2004}, and uses a Pedersen commitment scheme \cite{ped1991} to ensure verifiable secret sharing. The scheme allows any \texttt{M} validator nodes of the total \texttt{N} validator nodes on a blockchain to sign messages in a distributed way such that the private key shares do not need to be assembled to create a signature. Each validator node creates a signature share by signing the message using their private key share. Any \texttt{M} of the total \texttt{N} signature shares can be combined to create a valid signature. Importantly, the signature contains no information about which nodes signed, or what the threshold number of signatures (\texttt{M}) needed to create the signature is.

The Atomic Crosschain Transaction system uses BLS Threshold Signatures to prove that information came from a specific blockchain. For example, in Figure~\ref{fig:nested1}, nodes on \texttt{Private Blockchain B} can be certain of results returned by a node on \texttt{Private Blockchain C} for the function call to \texttt{funcC}, as the results are threshold signed by the validator nodes on \texttt{Private Blockchain C}. Similarly, validator nodes on \texttt{Private Blockchain A} can be certain that validator nodes on \texttt{Private Blockchain B} have mined the Subordinate Transaction, locked contract \texttt{ConB} and are holding the updated state as a provisional update because validator nodes sign a \textit{Subordinate Transaction Ready} message indicating that the Subordinate Transaction is ready to be committed.

\subsection{Multichain Nodes}
A Multichain Node is a logical grouping of one or more blockchain validator nodes, where each node is on a different blockchain. The blockchain nodes operate together to allow Crosschain Transactions. The Multichain Node on which the transaction is submitted must have Validator Nodes on all of the blockchains on which the Originating Transaction and Subordinate Transactions and Views take place. 

\begin{figure}
  \includegraphics[width=\linewidth]{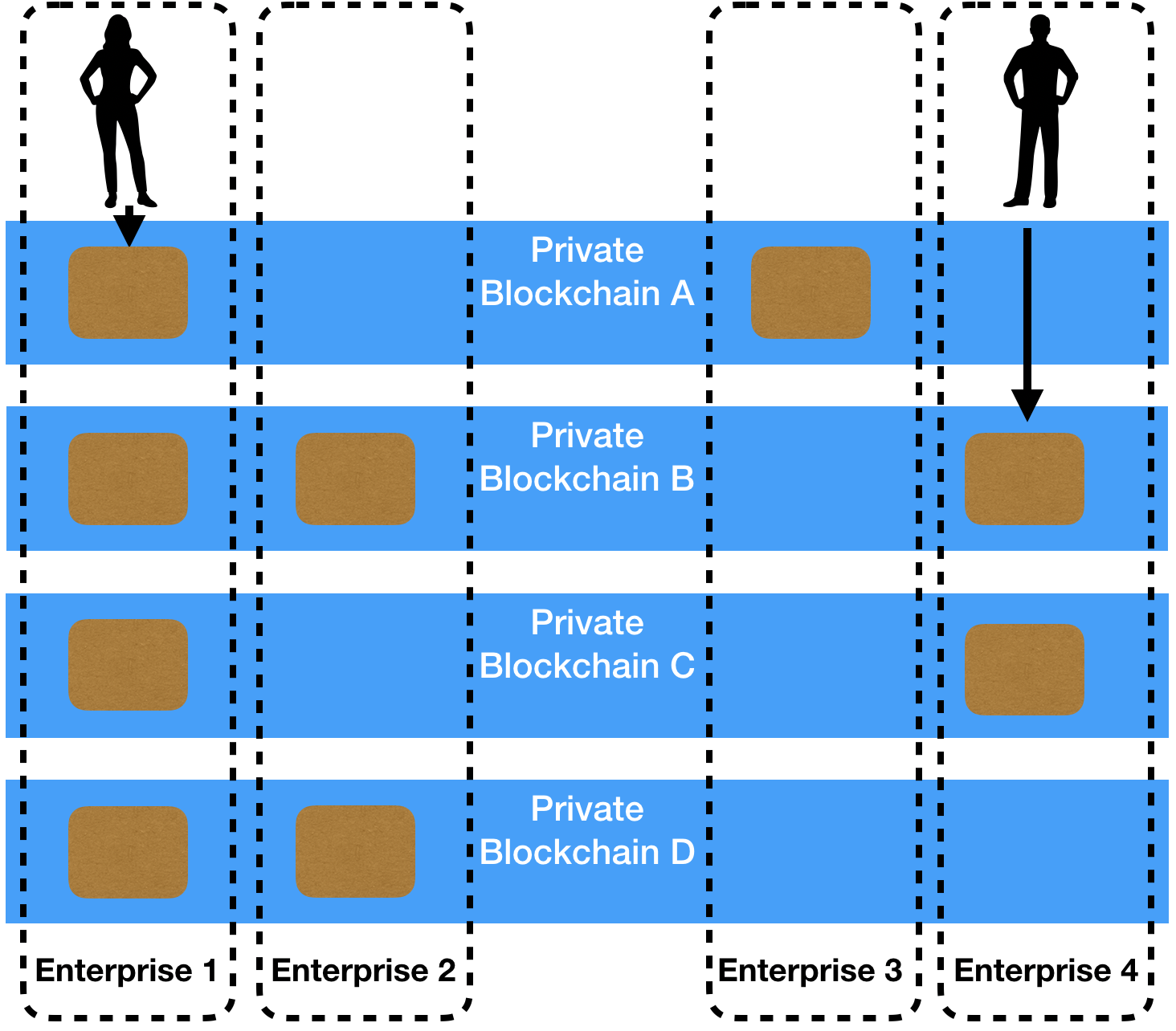}
  \caption{Multichain Nodes}
  \label{fig:multichain}
\end{figure}

Figure~\ref{fig:multichain} shows four enterprises that have validator nodes on \texttt{Private Blockchain A} to \texttt{Private Blockchain D}. An Enterprise 1 EOA can submit Atomic Crosschain Transactions that span \texttt{Private Blockchain A} to \texttt{Private Blockchain D} as Enterprise 1 has a Multichain Node that includes validator nodes on each blockchain. However, an Enterprise 4 EOA can only submit Atomic Crosschain Transactions that span \texttt{Private Blockchain B} and \texttt{Private Blockchain C} as Enterprise 4 only has validator nodes on \texttt{Private Blockchain B} and \texttt{Private Blockchain C}.

\subsection{Crosschain Coordination}
\textit{Crosschain Coordination Contracts} exist on \textit{Coordination Blockchains}. They allow validator nodes to determine whether the provisional state updates related to the Originating Transaction and Subordinate Transactions should be committed or discarded. The contract is also used to determine a common time-out for all blockchains, and as a repository of Blockchain Public Keys. 

When a user creates a Crosschain Transaction, they specify the Coordination Blockchain and Crosschain Coordination Contract to be used for the transaction, and the time-out for the transaction in terms of a block number on the Coordination Blockchain. The validator node that they submit the Originating Transaction to (the \textit{Originating Node}) works with other validator nodes on the blockchain to sign a \textit{Crosschain Transaction Start} message. This message is submitted to the Crosschain Coordination Contract to indicate to all nodes on all blockchains that the Crosschain Transaction has commenced. 

When the Originating Node has received \textit{Subordinate Transaction Ready} messages for all Subordinate Transactions, it works with other validator nodes to create a \textit{Crosschain Transaction Commit} message. This message is submitted to the Crosschain Coordination Contract to indicate to all nodes on all blockchains that the Crosschain Transaction has completed and all provisional updates should be committed. If an error is detected, then a \textit{Crosschain Transaction Ignore} message is created and submitted to the Crosschain Coordination Contract to indicate to all nodes on all blockchains that the Crosschain Transaction has failed and all provisional updates should be discarded. Similarly, if the transaction times-out, all provisional updates will be discarded.

\subsection{Contract Locking and Provisional State Updates}
When a contract is first deployed it is marked as a Lockable Contract or a Nonlockable Contract. A Nonlockable Contract, the default, is one which can not be locked. When a node attempts to update the state of a contract given an Originating or Subordinate Transaction, it checks whether the contract is \textit{Lockable} and whether it is \textit{locked}. The transaction fails if the contract is Nonlockable or if the contract is Lockable but is locked.

The act of mining an Originating Transaction or Subordinate Transaction and including it in a blockchain locks a contract. The contract can be unlocked when the Crosschain Coordination Contract is in the \textit{Committed} or \textit{Ignored} state, or when the block number on the Coordination Blockchain is greater than the Transaction Timeout Block Number. The Crosschain Coordination Contract will change from the \textit{Started} state to the \textit{Committed} state when a Crosschain Transaction Commit message is submitted to it, and it will change to the \textit{Ignored} state when a Crosschain Transaction Ignore message is submitted to it. 

When the Crosschain Coordination Contract indicates that the crosschain transaction has completed, or when the transaction has timed-out, Signalling Transactions are submitted on all blockchains that have locked contracts. The act of mining the Signalling Transaction unlocks all locked contracts.

\subsection{Crosschain Transaction Fields}
Originating Transactions, Subordinate Transactions, and Subordinate Views contain additional fields over and above those used in traditional Ethereum transactions \cite{crosschainwhitepaper}. The additional fields provide cross blockchain context and security.

\section{Experimental Setup}
\begin{figure} [h]     
\includegraphics[width=\columnwidth]{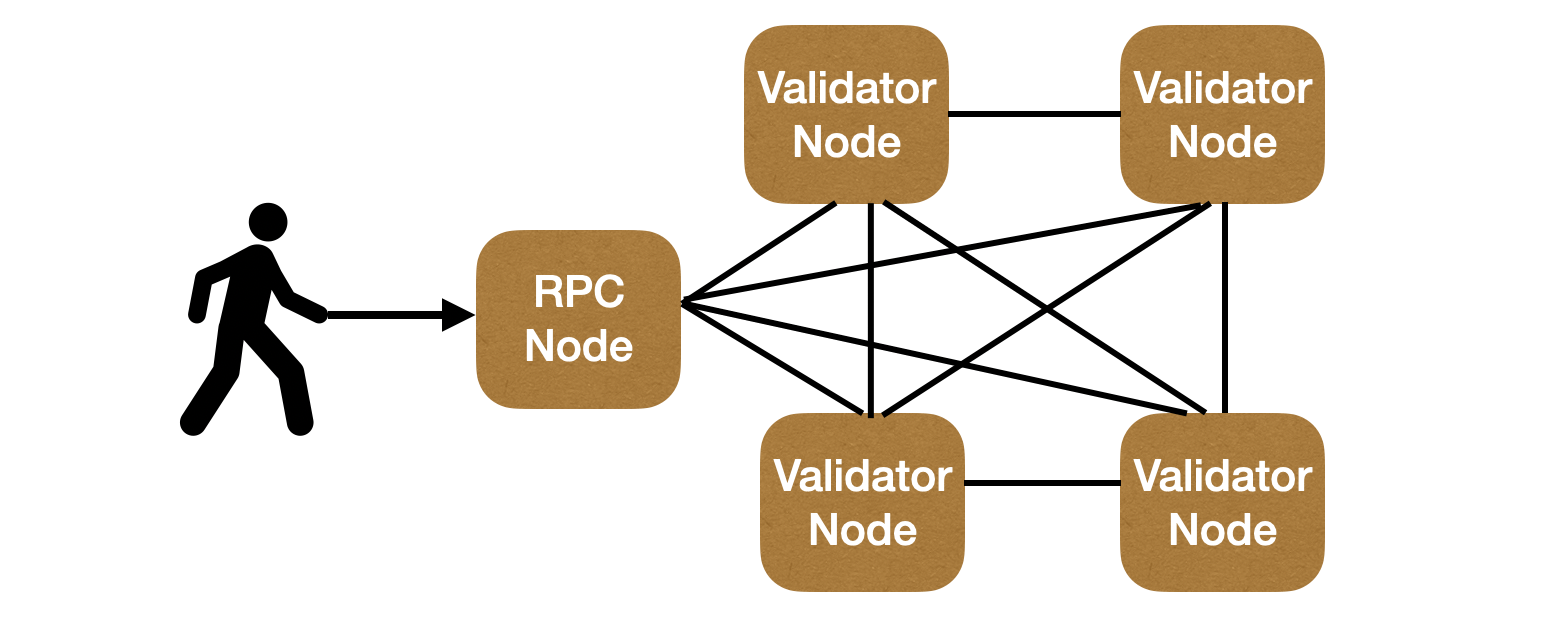}
\caption{Baseline Setup}
\label{fig:experiment}
\end{figure}

The performance of Hyperledger Besu version 1.4.4 with native crypto enabled, with an empty state trie, using IBFT2 consensus protocol, with no Byzantine nodes, when executing the \textit{open} function in the Hyperledger Caliper Benchmarks code Simple.sol\cite{caliper}, was measured using the configuration shown in Figure~\ref{fig:experiment}. All nodes were run on an Amazon Web Services c5d.4xlarge virtual machine (16 virtual CPUs, 32 GBytes of RAM, 450 GByte NVMe SSD, 2.25 Gbps elastic block storage bandwidth, 10 Gbps network connection). An EOA submitted transactions to a RPC Node that was connected with four validator nodes. All nodes were hosted in the same data centre.

In this configuration the limiting factor for performance was the processing power of the CPU. For the purposes of defining base line performance, it will be assumed that code for each scenarios executing in any function in a single blockchain transaction is able to execute at the measured rate of 375 transactions per second.



The Atomic Crosschain Transaction system as currently implemented involves two CPU intensive operations: completing a trial execution of transactions and verifying BLS threshold signatures. In a production implementation of the system the trial execution would be integrated into the first execution of a transaction within the consensus algorithm, thus removing that additional overhead. Verifying a BLS threshold signature for curve BN254 (formally known as BN128) was measured as 5 ms \cite{bls-perf}. 

\section{Results}
This section analyses the differences in requirements for operating a node given the scenarios described in Section~\ref{sec:scenarios} for using Atomic Crosschain Transactions as compared to executing separate non-atomic single blockchain transactions. Table~\ref{table:scenarios} succinctly describes the scenarios in terms of crosschain transactions components. In addition, for blockchains that have state updates, a Signalling Transaction must be submitted in addition to the Originating or Subordinate Transaction.

\begin{table}
  \centering
    \begin{tabular}{| l | l |}
    \hline
    Scenario & Description  \\
       \hline
       \hline
Hotel Train &  Originating Transaction with two Subordinate \\
                  &   Transactions.\\
       \hline
Supply Chain & Originating Transaction with one Subordinate \\
Provenance   & Transaction.\\
       \hline
Oracle           & Originating Transaction with Subordinate View. \\
       \hline
  \end{tabular}
  \caption{Analysis Scenarios}
  \label{table:scenarios}
\end{table}

Table~\ref{table:processing} shows the number of BLS signature verifications for certain nodes based on the type of transaction. Nodes that are the coordinating node on the originating blockchain for a transaction have to do more processing than any other node, having to generate the Start and Commit or Ignore messages and verify Subordinate Transaction Ready and Subordinate View Result messages. Other nodes on the originating blockchain only have to verify Subordinate Transaction Ready and Subordinate View Result messages. Table~\ref{table:processing2} shows the expected transaction rate given the number of BLS signature verifications and the fact that an additional Signalling Transaction is needed to commit or discard provisional updates on each blockchain.

\begin{table}
  \centering
    \begin{tabular}{| l | l |}
    \hline
    Transaction Type & Number of BLS Signature Verifications  \\
       \hline
       \hline
         \multicolumn{2}{|l|}{Originating Transaction} \\
       \hline
Coordinating node on   &  Two: One for Start and one for Commit or  \\
originating blockchain   &  Ignore messages during signature generation. \\
       \hline
Nodes on Coordination &  Two: One for Start and one for Commit or  \\
blockchain                     &  Ignore messages during submission to   \\
                                      &  Coordination Contract. \\
       \hline
       \hline
         \multicolumn{2}{|l|}{Subordinate Transaction} \\
       \hline
Nodes on originating         &  One to verify Subordinate Transaction \\
 blockchain    &  Ready message \\
       \hline
Coordinating node on    &  One for Subordinate Transaction Ready \\
subordinate blockchain  &  message during signature generation.\\
       \hline
       \hline
         \multicolumn{2}{|l|}{Subordinate View} \\
       \hline
Node on calling              &  One to verify Subordinate View Result \\
blockchain                   &  message. \\
       \hline
Coordinating node on    &  One for Subordinate View Result \\
subordinate blockchain  &  message during signature generation.\\
       \hline
  \end{tabular}
  \caption{BLS Signature Verifications}
  \label{table:processing}
\end{table}

\begin{table}
  \centering
    \begin{tabular}{| l | c | c |}
    \hline
    Scenario & \multicolumn{2}{c|}{Transactions Per Second}  \\
                   & \multicolumn{2}{c|}{Originating Blockchain}  \\
                   & Coordinating Node & Other Node  \\
       \hline
       \hline
Hotel Train &  39.5 & 65.2 \\
       \hline
Supply Chain & 49.2 & 96.8\\
Provenance   &   & \\
       \hline
Oracle           & 49.2 & 96.8 \\
       \hline
  \end{tabular}
  \caption{Transactions Per Second}
  \label{table:processing2}
\end{table}

For example, with the Hotel Train scenario, the transaction rate for the coordinating node on the originating blockchain is given by Equation~\ref{equation:tps}, where \texttt{Stx} and \texttt{Otx} are the Signalling and Originating Transactions that are assumed to execute at 375 tps and \texttt{TxR}, \texttt{St}, and \texttt{Co} are Transaction Ready, Start, and Commit messages, each assumed to cause 5ms overhead due to one BLS verify operation.

\begin{equation}
 \begin{aligned} Tx Rate & =   
  \cfrac{1}{Otx + Stx + 2 \times TxR + St + Co}\\
          & = \cfrac{1}{2 \times \cfrac{1}{375 tps} + 4 \times 5 ms} \\
          & = 39.5 tps
\end{aligned}          
\label{equation:tps}
\end{equation}

What Table~\ref{table:processing2} shows is that if all crosschain transactions are instigated from the same node, then the performance would be 39.5 tps for the Hotel-Train scenario. However, if the node instigating the crosschain transaction was distributed across nodes on the blockchain, then the performance will improve, approaching 65.2 tps because sometime a node will be acting as a coordinating node and at other times it will act as a non-coordinating node. 

\section{Effects of Byzantine Behaviour}
To generate a BLS threshold signature, the coordinating node on the multichain node that initiated the transaction sends the message to be signed to all validator nodes on the blockchain. Each validator node signs the message producing a signature share, and returns their signature share to the coordinating node. The coordinating node signs the message and combines a threshold number of signature shares using Lagrange Polynomials to produce the group signature. Prior to using the group signature the coordinating node needs to verify the signature. If the group signature is invalid, due to one or more of the validators supplying an invalid signature share, the coordinating node needs to verify each signature share individually to  determine the invalid signature shares, and then use additional signature shares to create a valid group signature. For subsequent signatures, the coordinating node would preferentially use signature shares from nodes it knows have historically produced valid signature shares. As such, though Byzantine nodes could temporarily cause a spike in required processing due to the use of Atomic Crosschain Transactions, they are unable to cause problems beyond one transaction.

\section{Conclusion}
This paper presents the first quantitative performance evaluation of Atomic Crosschain Transactions. These transactions provide synchronous inter-contract function calls across blockchains. No other technology provides similar atomic capability. They ensure updates across blockchains are either all committed or all ignored. This complex functionality comes with the cost of reduced blockchain system performance. For example, this technology reduces the performance of Hyperledger Besu from 375 tps to 39.5 tps if all transactions are instigated on the one node, or approaching 65.2 tps if the transactions are instigated on a variety of nodes, for the Hotel-Train scenario. 

The performance of Atomic Crosschain Transactions depends on the number of Subordinate Transactions and Views. Architects utilising this technology need to understand the call graph of their cross-blockchain function calls to determine the expected system performance. They need to determine important high value transactions that need to be cross-blockchain to minimise the impact on performance. 

Implementing Atomic Crosschain Transactions require complex changes to blockchain clients. These \textit{Layer 1} changes are not appropriate for scenarios in which customers are unwilling to have complex features added to blockchain clients. Given this, future research will focus on a \textit{Layer 2} approach to cross-blockchain transactions, leveraging many of the techniques developed for Atomic Crosschain Transactions.

\ifCLASSOPTIONcompsoc
  \section*{Acknowledgments}
\else
  \section*{Acknowledgment}
\fi
This research has been undertaken whilst I have been employed full-time at ConsenSys. I acknowledge the support of University of Queensland where I am completing my PhD, and in particular the support of my PhD supervisor Dr Marius Portmann. I acknowledge Dr Catherine Jones, Dr Raghavendra Ramesh, Horacio Mijail, Dr Marius Portmann, and Dr David Hyland-Wood for their thoughtful review comments and suggestions. I thank Ben Burns for running the Hyperledger Besu throughput tests.

\bibliographystyle{IEEEtran}
\bibliography{IEEEabrv,ref}

\end{document}